\documentclass[pra,aps,notitlepage,superscriptaddress,showpacs,showkeys]{revtex4-1}
\usepackage{amsmath,amsthm,amssymb,graphicx,subfigure,xcolor}
\usepackage[unicode=true]{hyperref}
\hypersetup{
     colorlinks=true,       		
     linkcolor=blue,          	
     citecolor=red,            
     urlcolor=magenta,           	
 }

\newtheorem*{theorem*}{Theorem}

\newtheorem*{corollary*}{Corollary}

\newtheorem*{lemma*}{Lemma}

\newtheorem*{proposition*}{Proposition}
\theoremstyle{definition}

\newtheorem*{definition*}{Definition}
\theoremstyle{remark}

\newtheorem*{remark*}{Remark}

\newcommand{\ket}[1]{|#1\rangle}

\newcommand{\ketbra}[2]{|#1\rangle\langle#2|}
\newcommand{\tr}{{\rm tr}}

\begin{document}

\title{A simple relation of guessing probability in quantum key distribution}

\author{Hong-Yi~Su}
\affiliation{College of Physics and Optoelectronic Engineering, Harbin Engineering University, Harbin 150001, People's Republic of China}
\affiliation{Graduate School of China Academy of Engineering Physics, Beijing 100193, People's Republic of China}

\date{\today}
\begin{abstract}
Given a communication system using quantum key distribution, the receiver can be seen as one who tries to guess the sender's information just as potential eavesdroppers do. The receiver-eavesdropper similarity thus implies a simple relation in terms of guessing probability and correctness of sifted keys, related with the distance-based, information-theoretic security. The tolerable regions of error rates determined by such a guessing-probability-based relation are shown to be close to those determined by security criteria. Thus, an alternative perspective on applying guessing probability in analyzing quantum key distribution issues is here provided. Examples of two specific protocols are illustrated. Our results contribute to evaluating an important element in communication study, and may provide useful reference for the security analysis of quantum key distribution protocols.
\end{abstract}

\maketitle

\section{Introduction}
Quantum key distribution (QKD) has been proved information-theoretically secure via quantum entropic arguments. For a communication (secrecy) system, however, as was stated in Claude Shannon's seminal paper, ``\emph{...[i]n general, the only difference between the decipherer's knowledge and the enemy cryptanalyst's knowledge is that the decipherer knows the particular key being used, while the cryptanalyst knows only the a priori probabilities of the various key in the set...}''~\cite{shannon49}. Despite such a difference, the receiver and the eavesdropper are rather similar in tasks like secret key distribution. This motivates us to evaluate eavesdropping using various measures other than the Shannon entropy or many of its variants in quantum theory.

Guessing probability~\cite{alimomeni12,issa17}, often requested by researchers alongside customers of cryptosystems, is among the candidates. It serves to estimate the eavesdropper's probability of correctly guessing the key transmitted from the sender to the receiver. It has drawn particular attention~\cite{konig09,yuen16,wang20} as quantum key distribution is being implemented in recent decades. There have been results pointing out that guessing probability of the key can provide an alternative perspective on QKD security. We for instance list a few relevant references as follows: (i) Guessing probability is associated with the min-entropy, in a way that the logarithm of the probability of guessing the sender's bits given eavesdropping provides a lower bound on the number of secret bits that can be extracted~\cite{konig09}. (ii) In Ref.~\cite{wang20}, researchers invoke finite-length analysis to take account of fluctuation effects and show that, after privacy amplification with the Toeplitz matrix, the guessing probability can be decreased by approximately $10^{-3268}$ of a previously obtained upper bound. (iii) Guessing probability depends on the specific protocols or key distillation post-processing~\cite{branciard05,bae07}. For instance, if the eavesdropper, in order to guess the sender's bit, takes the Helstrom strategy~\cite{helstrom76}, then the eavesdropping information will be directly determined by the guessing probability (see, e.g., Eq.~(27) in~\cite{branciard05} for an explicit relation between mutual entropy and guessing probablity). As shown in~\cite{bae07}, some key distillation post-processing can further decrease the eavesdropper's guessing probability. Using the classical advantage distillation~\cite{maurer93}, it is proved that (i.e., Eqs.~(71-73) in~\cite{bae07}) when security is compromised one can always have $F_N<P^{\rm success}_{\rm eq}$ (which corresponds to $P_B<P^\star_E$ in our notation defined below). Let us emphasize that the situation we shall consider is one without post-processing, thus differing from that in~\cite{bae07}. Additionally, study of guessing probability is nontrivial in itself, since it can be seen as a consequence of quantum state discrimination~\cite{bae-hwang-han11,bae-hwang13,bae13}, a fundamental task applicable in information processing. Note that the guessing probability considered here is computed in the protocols with trusted preparation and measurement; that is, the Hilbert dimensions of quantum states will have to be assumed fixed and known. Device-independent protocols, or semi-device-independent ones, which are without the assumption on dimensionality, are beyond our scope here (see, however, the discussion in the end of the paper).

That security of communication systems can be certified by using entropic criteria is attributed to Shannon, whose introduction of entropy into the communication studies led to the seminal formulation of information theory, by which an information-theoretic proof of security was for the first time attained for symmetric cryptography~\cite{shannon49}. In modern parlance, the security proof is motivated by a similarity between the receiver (Bob) who receives signals from the sender (Alice), and the eavesdropper (Eve) who may intercept and resend the signals transmitting in between. Hence a theoretical secrecy index---i.e., the equivocation---can be estimated by means of the conditional entropy $H(A|E)$, with $A$ representing the original message and $E$ the intercepted message. Much of the similarity between Bob and Eve is maintained in quantum cryptography~\cite{bb84,ekert91}. For instance, the quantum secure key rate is computed by a properly optimized difference between $S(A|E)$ and $S(A|B)$~\cite{ben-or05,renner-konig05,renner-phd}, where $S(x|y)$ denotes the quantum conditional von Neumann entropy resembling its classical counterpart. Nevertheless, quantum theory imposes a tradeoff on Bob and Eve, rendering Eve's intercepts detectable (see~\cite{gisin02,scarani09,lo2014,lo2016,pan2020} and references therein). It is noted that Eve, who presumably has an arbitrarily high-dimensional quantum memory, may be able to perform compatible measurements to discriminate Alice's outcomes under incompatible measurements~\cite{albert83}.

More specifically, the secure key rate formula can be derived from the finite-length, distance-based $\varepsilon$-security~\cite{ben-or05,renner-konig05,renner-phd}, and in the asymptotic limit it reduces to several well-known expressions, like $\mathcal{R}=1-2h(Q)$ for the Bennett-Brassard 1984 (BB84) prototol~\cite{bb84}, with $h(x):=-x\log_2 x-(1-x)\log_2(1-x)$ and $Q$ the error rate. For general composable protocols, the overall security level, $\varepsilon$, can be divided into two parts, the correctness $\varepsilon_{\rm cor}$ and the secrecy $\varepsilon_{\rm sec}$, subject to $\varepsilon_{\rm cor}+\varepsilon_{\rm sec}\leq\varepsilon$. Following the terminology and notations in~\cite{tomamichel12}, here the $\varepsilon_{\rm cor}$ bounds the probability that Alice and Bob fail to share identical keys of a certain length, namely, $P(A\neq B)\leq\varepsilon_{\rm cor}$, and the $\varepsilon_{\rm sec}$ bounds the trace-norm distance between a realistic Alice-Eve correlation and an ideal one, namely, $\frac{1}{2}\min_{\sigma_E}||\rho_{AE}-\tau_A\otimes\sigma_E ||_1\leq\varepsilon_{\rm sec}/(1-p_{\rm abort})$, where $p_{\rm abort}$ denotes the probability that the protocol is aborted, $\rho_{AE}$ is the quantum state representing the Alice-Eve correlation, $\tau_A$ denotes the fully mixed state, and $\sigma_E$ is an arbitrary state of Eve.

The correctness and secrecy are both essential requirements for a protocol to be secure, as can be seen by referring to the finite-length key rate formulas that usually involve both $\varepsilon_{\rm cor}$ and $\varepsilon_{\rm sec}$ (see, e.g., the formula in~\cite{tomamichel12}). In applications, it is reasonable to consider correctness and secrecy separately; however, the overall security level $\varepsilon$ makes more sense than its constituents. Hence $\varepsilon_{\rm cor}$ alone, while maybe very close to $\varepsilon$, does not imply security. Nevertheless, as we shall show in this paper, study focused on $\varepsilon_{\rm cor}$ bears a merit that it has potential to intuitively connect the Alice-Bob correlation with the guessing probability, i.e., with the Alice-Eve correlation. The unobservable Alice-Eve correlation can thus be evaluated via the observable Alice-Bob correlation. By treating Bob and Eve from a similar perspective, i.e., both can be seen as trying to guess keys from Alice, a guessing-probability-based relation can be established, giving rather efficient predictions as by the entropic security criterion.

Before presenting the setup and formulation of guessing probability, we would like here to discuss the guessing-probability-based relation and the security criterion. On the one hand, since $\varepsilon_{\rm cor}$ is no greater than $\varepsilon$ in magnitude, an ``$\varepsilon_{\rm cor}$-secure'' protocol must be considered more secure than an $\varepsilon$-secure protocol. For this reason the guessing-probability-based relation, motivated by $\varepsilon_{\rm cor}$, is likely to impose slightly stricter requirement than the entropic criterion, motivated by $\varepsilon$, upon some essential factors, e.g., critical error rates. This is the case in our results shown in the next section, particularly in the two examples, to which we refer the readers for detail. It is stressed, however, that the gap between the different predictions should not be interpreted as some overestimate by the established security criterion, but rather as immediate consequences of the $\varepsilon$-security formalism. On the other hand, results so far in existing literature, concerning guessing probabilities that match the hierarchy between $\varepsilon_{\rm cor}$ and $\varepsilon$, have rarely been developed. In other words, a desired $\varepsilon_{\rm cor}$-motivated relation ought to exist in order to impose requirements stricter than---if not equivalent to---the $\varepsilon$-security criteria upon QKD protocols. Will it be done, for instance, with any carefully built relations, which could involve keys before the final phase but still remain intuitively simple? This question raises our principal motivation of quantifying the guessing probability, from which we are then able to construct meaningful relations and give an affirmative answer to the question.

\section{Quantifying guessing probability}
\subsection{Brief description of the setup}
To build such a relation, one will first have to evaluate the guessing probability via some observables (usually the error rates, known to Alice and Bob). The evaluation requires an optimization procedure, because the Alice-Bob density matrix, representing the Alice-Bob correlation, can be reduced from non-unique Alice-Bob-Eve purification states; namely, different purifications may yield different guessing probabilities for a same Alice-Bob correlation. Then, one can have a guessing-probability-based relation in terms of the Alice-Bob correlation and the Alice-Eve correlation. A desired, sufficiently efficient relation should impose some stricter requirements on the composable QKD protocols, in accord with the hierarchy between $\varepsilon_{\rm cor}$ and $\varepsilon$.

Let us briefly describe the setup. We consider the QKD protocols in which Alice sends Bob keys which Eve may intercept and resend. Instead of the final key, here we are concerned to study the sifted key---i.e., the key after the sifting phase and before the classical postprocesses such as information reconciliation and privacy amplification. We consider collective attacks~\cite{biham97,biham05} so that it suffices to study a single run of preparation and measurement with the quantum state of Alice and Bob. Let $\mathcal{H}_A(d)$, $\mathcal{H}_B(d)$ and $\mathcal{H}_E(td)$ be the Hilbert spaces of Alice, Bob and Eve, respectively. Alice has $t$ sets of $d$-dimensional bases in which to measure, hence the dimension $td$ of Eve. The guessing probability then reads $P_E=\sum_{i=0}^{t-1} P(e= a_i+i\times d)$, $a_i$ and $e$ being the outcomes of events of Alice and Eve. Motivated by $\varepsilon_{\rm cor}$, also in the light of Shannon's idea on the similarity between Bob and Eve, we present an equally significant quantity $P_B=\sum_{i=0}^{t-1}\wp_i P(a_i=b_i)$, with $b_i$'s similarly denoting the outcomes of events of Bob, $\wp_i$ the probability that Alice chooses the $i$-th basis to perform measurements, and $\sum_i\wp_i=1$.

Given transmission through a noisy quantum channel, Alice prepares and sends out a sequence of qubits, which are then received and measured by Bob, and which may also be intercepted by Eve and resent to Bob. After measurements and the sifting phase, Bob gets a sequence of bits, along with a $P_B$ by comparing some of the bits in public with Alice, and Eve gets a result of Alice's bits, along with a $P_E^\star$ by listening to the publicized bases in the sifting phase and optimizing her strategies. Here $P_E^\star$ represents the optimized $P_E$. The Bob-versus-Eve scenario allows us to propose the relation $P_B>P_E^\star$ as an index that delivers nontrivial indications of the error rates in QKD.

The justification is as follows. For a certain bit of Alice, it could be that Bob's guess is incorrect (i.e., an error appears) but Eve's guess is correct. This is not the worst case for key generation, however. For fixed $P_B$ and $P_E^\star$, the worst case is that for any of Alice's bits, if Bob's guess is incorrect, then Eve's guess is incorrect, too. Eve can thus make the most of her correct guesses, as it is Bob's correct guesses that are used to generate identical keys with Alice subsequently.

In what follows, we will first present a definition of the maximum guessing probability and apply it to three generic QKD protocols. Then, we will explicitly derive a general key-rate formula in the information-theoretic manner. As examples, we will consider two well-known protocols and compute the maximum guessing probabilities. The determinations of the critical error rates by the guessing-probability-based relations and by the entropic security criteria will be compared and discussed.

\subsection{Definition of the maximum guessing probability}
To begin with, it is convenient to consider the entanglement-based scenario~\cite{bbm92,bennett96}. The general state shared by Alice and Bob reads
$\rho_{AB}=R R^\dagger$, where $R=U\sqrt{\Lambda}V^\dagger$,
$R$ and $U$ are $d^2\times d^2$ matrices, $U$ denotes the unitary transform that diagonalizes $\rho_{AB}$, $\Lambda$ is a $d^2\times td$ matrix, with $\Lambda_{k}$ representing the eigenvalues of $\rho_{AB}$, and $V$ is an arbitrary $td\times td$ unitary representation of the $SU(td)$ group. The entries of $R$ can be used as coefficients of any purification of $\rho_{AB}$, namely,
$\ket{\psi}_{ABE}=\sum_{i,j}R_{ij}\ket{i}_{AB}\ket{j}_E$,
with $\ket{i}_{AB}\in\mathcal{H}(d^2)$ and $\ket{j}_E\in\mathcal{H}(td)$~\cite{su20}.
Because of the Schmidt decomposition, we define $\sum_i U_{ik}\ket{i}_{AB}:=\ket{k_U}_{AB}$ and $\sum_j(V^\dagger)_{kj}\ket{j}_E:=\ket{k_V}_E$, and the purification state immediately becomes $\ket{\psi}_{ABE}=\sum_k\sqrt{\Lambda_k}\ket{k_U}_{AB}\ket{k_V}_E$, with $k=0,~1,\cdots,~\min\{d^2-1,~td-1\}$.
Here, the basis $\{\ket{i}_{AB}\}$ can be chosen in such a way that $\ket{k_U}_{AB}=\ket{\phi_k}_{AB}$, where $\ket{\phi_k}_{AB}$ are the bases for the diagonalized $\rho_{AB}=\sum_k\Lambda_k\ketbra{\phi_k}{\phi_k}_{AB}$, and the basis $\{\ket{j}_E\}$ can be chosen as the computational basis such that $\ket{k_V}_E=(V^\dagger)^{\rm T}\ket{k}_E$.
The guessing probability, depending clearly on Eve's strategy $V$, then equals
\begin{align}
   &P_E=\tr \biggr[\Bigr(\sum_{i=0}^{t-1}\sum_{a_i=0}^{d-1}\bigr|a_i\bigr\rangle\bigr\langle a_i\bigr|_A\otimes\bigr|e\bigr\rangle\bigr\langle e\bigr|_E \Bigr)\bigr|\psi\bigr\rangle\bigr\langle \psi\bigr|_{ABE}\biggr],\nonumber\\
   &{\rm subject~to}~~ e=a_i+i\times d.\nonumber
\end{align}
One who studies security issues must take into account the worst circumstances. The quantity that indeed makes sense in evaluating Eve's guesses must be the maximum guessing probability,
\begin{equation}
  P^\star_E=\max_V P_E,
\end{equation}
where the optimization is accomplished by running all $V\in SU(td)$, along with nonnegative $\Lambda_i$'s subject to observables of Alice and Bob.

Here, we remark on the $V$ which corresponds to Eve's strategy to guess Alice's measurement results. For each $\rho_{AB}$ shared by Alice and Bob, Eve can be presumed to hold a purification such that the complete state is described by the purification state $\ket{\psi}_{ABE}$. The purification is not unique, and $V$ is the unitary transformation that links these purification states which reduce to the same $\rho_{AB}$. The importance of numerating all possible $V$ can then be seen in an example with the BB84 protocol presented in Sec.~\ref{bb84}, particularly the discussion around Eqs.~(\ref{purification_example}) through (\ref{purification_worst_case}).

\subsubsection{\label{symmetry}Some symmetric properties}
A symmetry exists in the correlations of measurements in typical QKD protocols in which qubits (i.e., $d=2$) are used and Alice and Bob share the Bell-diagonal state~\cite{kraus05,renner05}, namely, $\ket{\phi_0}=(\ket{00}+\ket{11})/\sqrt{2}$, $\ket{\phi_1}=(\ket{00}-\ket{11})/\sqrt{2}$, $\ket{\phi_2}=(\ket{01}+\ket{10})/\sqrt{2}$, and $\ket{\phi_3}=(\ket{01}-\ket{10})/\sqrt{2}$.
To see it, let us first take $\mathcal{U}=e^{-i\frac{\theta}{2}\sigma_{\varphi}}$ for Alice, with $\sigma_\varphi=\sigma_y\cos\varphi-\sigma_x\sin\varphi$. The unitary is to transform a projector along the $\hat z$ direction to one along an arbitrary $\hat n$ direction.
It is simple to verify that $\mathcal{U}\sigma_z\mathcal{U}^\dagger=\hat n\cdot\vec\sigma$, and that
$\mathcal{U}\ket{0}=\ket{+n}$, $\mathcal{U}\ket{1}=-e^{-i\varphi}\ket{-n}$,
where $\ket{\pm n}$ are a set of orthogonal normalized states along $\hat n$, which by parametrization read $\ket{+n}:=\ket{\theta,\varphi}=\cos\frac{\theta}{2}\ket{0}+\sin\frac{\theta}{2}e^{i\varphi}\ket{1}$ and $\ket{-n}:=\ket{\overline{\theta,\varphi}}=\sin\frac{\theta}{2}\ket{0}-\cos\frac{\theta}{2}e^{i\varphi}\ket{1}$.
Likewise, for Bob, a set of orthogonal normalized states along $\hat n'$ can be written as $\ket{+n'}:=\ket{\theta,-\varphi}$ and $\ket{-n'}:=\ket{\overline{\theta,-\varphi}}$.
Then, for the correlations
\begin{equation}
  P_{\pm n,\pm n'}=\tr \Bigr[\ketbra{\pm n}{\pm n}_A\otimes\ketbra{\pm n'}{\pm n'}_B~\rho_{AB}\Bigr],
\end{equation}
the following relations hold:
\begin{align}
  {P}_{+n,+n'}={P}_{-n,-n'}, ~~~~{P}_{+n,-n'}={P}_{-n,+n'}.
\end{align}
Because Alice (Bob) has $t$ directions to measure along, with probability $\wp_i$ for choosing each and subject to $\sum_{i=0}^{t-1} \wp_i=1$, the $P_{\pm n,\pm n'}$ when computed from the data pool have to be rescaled by multiplying $\wp_i$. We thus for each $i$ have specifically
\begin{equation}\label{p-01}
\begin{split}
  &{P}_{+n_i,+n'_i}={P}_{-n_i,-n'_i}
  =\frac{\wp_i}{2}\times\Delta_i,\\
  &{P}_{+n_i,-n'_i}={P}_{-n_i,+n'_i}
  =\frac{\wp_i}{2}\times(1-\Delta_i),
\end{split}
\end{equation}
with $\Delta_i=\Lambda_0+\Lambda_1\cos^2\theta_i+\Lambda_2\sin^2\theta_i\cos^2\varphi_i+\Lambda_3\sin^2\theta_i\sin^2\varphi_i$.
The reason we choose $\hat n$ and $\hat n'$ in demonstrating the symmetry is that it is $\ket{\phi_0}$ that we take as the maximally entangled state in the ideal quantum channel. Clearly, if we take any of other Bell states as the maximally entangled state in the ideal quantum channel, a corresponding $\hat n''$ direction may be found and used for Bob to demonstrate with Alice the symmetry of $P_{\pm n,\pm n''}$.

\subsubsection{Typical types of QKD protocols}
The $\Lambda_i$'s are not all independent of one another, since they are connected with observables, i.e., the error rates, which are known to Alice and Bob.
For each $i$, due to (\ref{p-01}) and
\begin{equation}
  \varepsilon_{i}:=\Bigr[{P_{+n_i,-n'_i}+P_{-n_i,+n'_i}}\Bigr]\biggr/\sum_{m_i, m'_i} P_{m_i,m'_i},
\end{equation}
(which corresponds to the correctness of the sifted key and should not be confused with the overall security level $\varepsilon$ for the final key) the sum being over $m_i=\pm n_i$ and $m'_i=\pm n'_i$, we find $\varepsilon_{i}=1-\Delta_i$. Without loss of generality,
let us hereafter take $\varepsilon_0:=\varepsilon_{\hat z}=\Lambda_2+\Lambda_3$, i.e., $\theta_0=\varphi_0=0$.
Then we can list three generic protocols:

\emph{Protocol I:} A four-state protocol with $t=2$ measuring directions $\hat n_{0,1}$. There are two error rates relating to the $\Lambda_i$'s, so only one parameter in $\rho_{AB}$, say $\lambda_3$, is free. That is, $\Lambda_0=1-(\cos^2\varphi_1-\cot^2\theta_1)\varepsilon_0-(1/\sin^2\theta_1)\varepsilon_1+\Lambda_3\cos2\varphi_1$, $\Lambda_1=1-\varepsilon_0-\Lambda_0$, and $\Lambda_2=\varepsilon_0-\Lambda_3$. Hence, the $P^\star_E$ with respect to $\varepsilon_{0,1}$ can be computed by numerating $V\in SU(4)$ and $\Lambda_3$, subject to $0\leq\Lambda_i\leq1$ for any $i$.

\emph{Protocol II:} A six-state protocol with $t=3$ measuring directions $\hat n_{0,1,2}$. Three error rates are related to the $\Lambda_i$'s, so all of them are fixed. The $P^\star_E$ with respect to $\varepsilon_{0,1,2}$ can be computed by numerating $V\in SU(6)$.

\emph{Protocol III:} A $2t$-state protocol with $t>3$ measuring directions $\hat n_{0,1,...,t-1}$. The first three error rates are connected with $\Lambda_i$'s, and the remaining error rates are determined by the first three as
\begin{equation}\label{determine}
\begin{split}
  &\varepsilon_{k}=(1+\delta_1-\delta_2)\varepsilon_{0}-\delta_1\varepsilon_{1}+\delta_2\varepsilon_{2},
\end{split}
\end{equation}
for $3\leq k<t$, with
\begin{equation}
\begin{split}
  &\delta_1=\frac{\sin^2\theta_k\sin(\varphi_2-\varphi_k)\sin(\varphi_2+\varphi_k)}{\sin^2\theta_1\sin(\varphi_1-\varphi_2)\sin(\varphi_1+\varphi_2)},\\
  &\delta_2=\frac{\sin^2\theta_k\sin(\varphi_1-\varphi_k)\sin(\varphi_1+\varphi_k)}{\sin^2\theta_2\sin(\varphi_1-\varphi_2)\sin(\varphi_1+\varphi_2)}.
\end{split}
\end{equation}
The $P^\star_E$ with respect to $\varepsilon_{0,1,...,t-1}$ can be computed by numerating $V\in SU(2t)$.

\subsection{Derivation of a general secure key rate}
We consider the information-theoretic security of QKD, for which the secure key rate is computed by $\mathcal{R}=\max\{\mathcal{I}_{AB}-\max_{\Lambda_i}\chi_{AE},~0\}$~\cite{shor2000,winter05}, with the optimization of the Holevo quantity $\chi_{AE}$~\cite{holevo73} done by running nonnegative independent $\Lambda_i$'s. Given the key is generated in all bases of the measurements, in what follows we explicitly derive the key rate. It is noted that the entropic security criterion has been well established and the derivation below is not an original contribution of this paper.

To the end, let us first derive the mutual information $I_{AB}=\sum_{i,j}\sum_{m_i,m'_j} P_{m_i,m'_j}\log_2 P_{m_i,m'_j}/P_{m_i} P_{m'_j}$, where $P_{m_i}$ and $P_{m'_j}$, with $m_i=\pm n_i$ and $m'_j=\pm n'_j$, denote marginals for Alice and for Bob, respectively. We consider the protocols where cross-term data is not used; i.e., $P_{m_i,m'_j}$=0 for $i\neq j$, such that the mutual information can be computed independently for each $i$. With the symmetry in (\ref{p-01}), we find then $I_i=\wp_i(1-h(\varepsilon_{i}))-\wp_i\log_2\wp_i$, a sum of which yields
\begin{equation}
  I_{AB}=1-\sum_i \wp_i h(\varepsilon_{i})+H(\wp_i).
\end{equation}
The Shannon entropy, $H(\wp_i)=-\sum_i\wp_i\log_2\wp_i$, represents the information of the choices of bases, and it is irrelevant for the key generation. Hence the mutual information we actually use is one with this quantity deducted; namely,
\begin{equation}
  \mathcal{I}_{AB}=I_{AB}-H(\wp_i).
\end{equation}

Let us next consider the eavesdropping in which Eve holds states $\rho_{E|\pm n_i}$ conditional on Alice's measurements with probability $\wp_i$ for each $i$. The Holevo quantity for our purpose here reads $\chi_{AE}=S(\rho_{AB})-\sum_{i}\wp_i\sum_{m_i} p_{m_i} ~S(\rho_{E|m_i})$, where $p_{m_i}$ with $m_i=\pm n_i$ denotes the probability that Alice measures her qubit with projector $\ketbra{\pm n_i}{\pm n_i}$, and we have taken the identity $S(\rho_E)=S(\rho_{AB})$ as $\ket{\psi}_{ABE}$ is pure.
The conditional states $\rho_{E|\pm n_i}$ are of rank two, and have eigenvalues
$\lambda_{+n_i}^\pm=\lambda_{-n_i}^\pm=[1\pm\sqrt{\xi_i+\eta_i}]/2$, where $\xi_i=(\mu_+ - \nu_+)^2\cos^2\theta_i$ and $\eta_i=(\mu_-^2+\nu_-^2+2\mu_- \nu_- \cos2\varphi_i)\sin^2\theta_i$, with $\mu_\pm=\Lambda_0\pm\Lambda_1$ and $\nu_\pm=\Lambda_2\pm\Lambda_3$. (These eigenvalues were computed in the $\hat x\hat z$-plane in~\cite{acin09}; see also~\cite{su20}.)
It is straightforward to have $S(\rho_{AB})=H(\Lambda_i)$. The Holevo quantity is computed then
\begin{equation}
  \chi_{AE}=H(\Lambda_i)+\sum_{i,m_i,q}\wp_i p_{m_i}\lambda_{m_i}^q\log_2\lambda_{m_i}^q,
\end{equation}
with the sum over $q=\pm$, $m_i=\pm n_i$, and $i=0,...,t-1$.
Given the symmetry of the Bell-diagonal state, it holds that $p_{+n_i}=p_{-n_i}=1/2$ for any $\hat n_i$.

It is remarked that for the $t=2$ protocols, the relations between $\varepsilon_k$'s and $\Lambda_i$'s serve as constraints in optimizing; and for the $t\geq3$ protocols, there is no need to optimize with $\Lambda_i$'s but the $\varepsilon_k$'s must satisfy the relations in Protocol III. Obviously, under proper circumstances the general criterion reduces to some established formulas, like $\mathcal{R}=1-2h(\varepsilon_{\hat z})$ and $1-h(3\varepsilon_{\hat z}/2)-(3\varepsilon_{\hat z}/2)\log_2 3$, which have already been obtained in \cite{shor2000,lo01} for the original BB84 and six-state protocols, respectively.

\subsection{\label{bb84}Examples of guessing probability}
\subsubsection{The BB84 protocol}
\begin{figure}[tb]
\subfigure[]{\includegraphics[width=40mm]{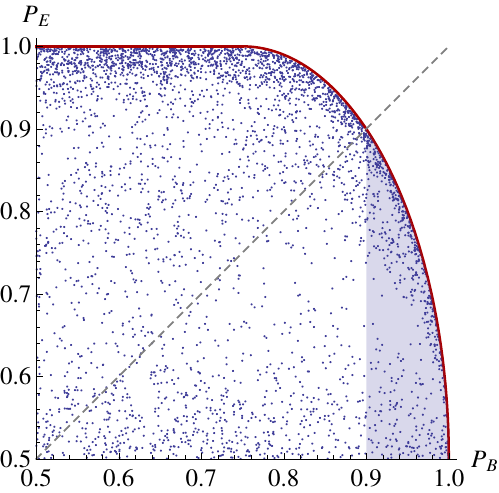}}\hspace{4mm}
\subfigure[]{\includegraphics[width=40mm]{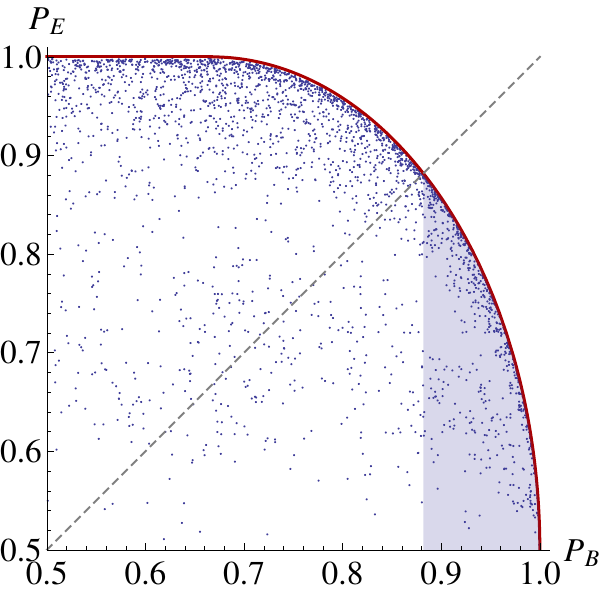}}
  \caption{(Color online) The $P_B$-versus-$P_E$ relations for the BB84 (left) and six-state (right) protocols, where $P_B=1-\varepsilon$, and the upper bounds (red curves) represent the $P^\star_E$'s in (\ref{pe-four}) and (\ref{pe-six}). The data (blue points) were collected with around 3800 (left) and 2750 (right) Bell-diagonal states generated by random $\Lambda_i$'s, alongside random irreducible unitary representations of $SU(4)$ (left) and $SU(6)$ (right) groups, respectively. The dashed lines correspond to $P_B=P_E$ so the points falling in shaded regions all have the property that Eve's maximum guessing probability is less than $P_B$. Note, however, that because Eve has less limits on Hilbert-space dimensions than Bob, $P_E^\star$ reaches 1 well before $P_B$ decreases to 1/2 (see, e.g., points $(3/4,1)$ and $(2/3,1)$ on the left and right red curves, respectively).  }\label{fig1}
\end{figure}

The protocol belongs to the four-state protocol (i.e., Protocol I) with $\theta_0=\varphi_0=0$, $\theta_1=\pi/2$, $\varphi_1=0$,
and $\wp_0=\wp_1=1/2$. Let us present an optimal quantum state $\Lambda_0=(1+\kappa)/2$, $\Lambda_1=\Lambda_2=(1-\kappa)/4$, and $\Lambda_3=0$, along with an optimal unitary transform $V\in SU(4)$ such that
\begin{equation}\label{optimum_V}
\begin{split}
  \ket{0_V}_E&=\frac{1}{2}\ket{0}_E-\frac{1}{2}\ket{1}_E+\frac{1}{2}\ket{2}_E-\frac{1}{2}\ket{3}_E,\\
  \ket{1_V}_E&=\frac{1}{\sqrt{2}}\ket{0}_E+\frac{1}{\sqrt{2}}\ket{1}_E,\\
  \ket{2_V}_E&=\frac{1}{\sqrt{2}}\ket{2}_E+\frac{1}{\sqrt{2}}\ket{3}_E,\\
  \ket{3_V}_E&=\frac{1}{2}\ket{0}_E-\frac{1}{2}\ket{1}_E-\frac{1}{2}\ket{2}_E+\frac{1}{2}\ket{3}_E,
\end{split}
\end{equation}
where $\ket{j}_E$ takes the computational basis as previously stated. The error rates equal to $\varepsilon_0=\varepsilon_1=(1-\kappa)/4:=\varepsilon\in[0,~1/4]$ and the purification $\ket{\psi}_{ABE}=\sum_{k=0}^3 \sqrt{\Lambda_k}\ket{\phi_k}_{AB}\ket{k_V}_E$ are henceforth obtained. The maximum guessing probability then equals
\begin{align}
  P^\star_E&=\sum_{j=0,1}\tr\biggr[\biggr(\Bigr|+n_j\Bigr\rangle\Bigr\langle +n_j\Bigr|_A\otimes\Bigr|2j\Bigr\rangle\Bigr\langle 2j\Bigr|_E
  +\nonumber\\
  &~~~~\Bigr|-n_j\Bigr\rangle\Bigr\langle -n_j\Bigr|_A\otimes\Bigr|2j+1\Bigr\rangle\Bigr\langle 2j+1\Bigr|_E \biggr)\Bigr|\psi\Bigr\rangle\Bigr\langle\psi\Bigr|_{ABE}\biggr]\nonumber  \\
  &=\frac{1}{2}+\sqrt{\frac{\Lambda_0}{2}}(\sqrt{\Lambda_1}+\sqrt{\Lambda_2})\nonumber\\
  &=\frac{1}{2}+\sqrt{2\varepsilon(1-2\varepsilon)}.\label{pe-four}
\end{align}
It equals unity for $\varepsilon\in(1/4,~1/2]$.
We plot the $P_B$-versus-$P_E$ figure (see Fig.~\ref{fig1}(a)) for arbitrary Bell-diagonal states, confirming the maximum of (\ref{pe-four}).

Let us come to see in detail how $V$ relates to Eve's strategy. (Of course, Eve's strategy also involves the choices of $\Lambda_i$'s; namely, it is possible to take other $\Lambda_k$'s to write a different $\rho_{AB}$ that yield a same $\varepsilon$, but here we take it for simplicity and only focus on $V$.) If Eve uses an strategy that $V=\openone$, we have $\ket{k_V}_E=\ket{k}_E$, and
\begin{equation}\label{purification_example}
  \begin{split}
    \ket{\psi}_{ABE}&=\sqrt{1-2\varepsilon}\ket{\phi_0}_{AB}\ket{0}_E+\sqrt{\varepsilon}\ket{\phi_1}_{AB}\ket{1}_E\\
    &~~~+\sqrt{\varepsilon}\ket{\phi_2}_{AB}\ket{2}_E.
  \end{split}
\end{equation}
It is obvious to see that Eve cannot discern Alice's qubits by her projective measurements $\ketbra{k}{k}_E$, because for each component in (\ref{purification_example}) with respect to $k$, Alice's state is fully random.
Accordingly, we find
\begin{align}
  P_E&=\tr\bigr[ (\ketbra{0}{0}_A\otimes\ketbra{0}{0}_E +\ketbra{1}{1}_A\otimes\ketbra{1}{1}_E+ \nonumber\\
   &~~~\ketbra{+}{+}_A\otimes\ketbra{2}{2}_E +\ketbra{-}{-}_A\otimes\ketbra{3}{3}_E)\rho_{ABE} \bigr]\nonumber\\
   &=1/2,\label{non-optimal-pe}
\end{align}
which is smaller than (\ref{pe-four}).
Hence, $V=\openone$ is not an optimal strategy for Eve. By ``optimal strategy,'' we mean that the $V$, for instance the one that induces (\ref{optimum_V}), can enable Eve to guess Alice's measurement results with the largest probability, under her projective measurements $\ketbra{k}{k}_E$. In other words, Eve must try to find an optimal $V$ so that with a purification $\ket{\psi}_{ABE}=\sum_{k=0}^3 \sqrt{\Lambda_k}\ket{\phi_k}_{AB}\ket{k_V}_E$ she can obtain the maximum of $P_E$, i.e., the $P^\star_E$, under the projection $\sum_k \ketbra{n_k}{n_k}_A\otimes\ketbra{k}{k}_E$, with $\ket{n_0}_A=\ket{0}_A,~\ket{n_1}_A=\ket{1}_A,~\ket{n_2}_A=\ket{+}_A$, and $\ket{n_3}_A=\ket{-}_A$.

In the ideal case with $\varepsilon=0$, Eve does not have any optimal strategy $V$ because the purification state now reads
\begin{equation}
\begin{split}
  \ket{\psi}_{ABE}=\ket{\phi_0}_{AB}\ket{0_V}_E,
\end{split}
\end{equation}
from which we always get $P^\star_E=1/2$. In the worst case with $\varepsilon=1/4$ and letting Eve choose the optimal strategy (\ref{optimum_V}), the purification state reads
\begin{equation}\label{purification_worst_case}
  \begin{split}
    \ket{\psi}_{ABE}&=\frac{1}{\sqrt{2}}\ket{\phi_0}_{AB}\ket{0_V}_E+\frac{1}{2}\ket{\phi_1}_{AB}\ket{1_V}_E\\
    &~~~~~+\frac{1}{2}\ket{\phi_2}_{AB}\ket{2_V}_E\\
    &=\frac{1}{2}\Bigr[ \Bigr|00\Bigr\rangle_{AB}\Bigr|0\Bigr\rangle_E- \Bigr|11\Bigr\rangle_{AB}\Bigr|1\Bigr\rangle_E \\
    &~~~~~+\Bigr|++\Bigr\rangle_{AB}\Bigr|2\Bigr\rangle_E-\Bigr|--\Bigr\rangle_{AB}\Bigr|3\Bigr\rangle_E \Bigr],
  \end{split}
\end{equation}
from which we find $P^\star_E=1$, as it should be.

Let $P_B>P^\star_E$, and we find $\varepsilon<10\%:=\varepsilon_{\rm cr}$. The bound is close to that determined by the entropic criterion, in the sense that the same error rate yields $\mathcal{R}=1-2h(\varepsilon_{\rm cr})\simeq0.062$. Note that the information of the basis is $\log_2 2=1$ bit, which has been deducted from the $\mathcal{R}$ here. Meanwhile, let $\mathcal{R}=0$ then we find $\tilde\varepsilon_{\rm cr}\simeq11\%$, yielding $P_B\simeq89\%$, which is slightly less than $P^\star_E\simeq91\%$.

A more general comparison is presented in Table~\ref{table}. It is noted that the critical value of the error rate for $\mathcal{R}=0$ varies with the measuring directions. The critical value implied from $P_B=P_E^\star$ varies with the directions, too. These variations with respect to $\varphi_1$ are due to that it is $\ket{\phi_0}$ that we take as the maximally entangled state in the ideal case. For each $\varphi_1$, if we take $\ket{\phi'_0}=e^{i\varphi_1\sigma_z}\otimes e^{i\varphi_1\sigma_z}\ket{\phi_0}$ as the maximally entangled state in the ideal case, i.e., the coordinate is rotated along the $\hat z$ axis such that the $\hat n_1$ direction becomes a new $\hat x$ axis, then there will be no variations, and the critical value from $P_B=P_E^\star$ will again be slightly smaller than that from $\mathcal{R}=0$, as in the $\varphi_1=0$ case.

\subsubsection{The six-state protocol}
The protocol belongs to Protocol II with $\theta_0=\varphi_0=0$, $\theta_1=\pi/2$, $\varphi_1=0$, $\theta_2=\varphi_2=\pi/2$,
and $\wp_0=\wp_1=\wp_2=1/3$. Likewise, we present an optimal quantum state $\Lambda_0=(1+\kappa)/2$ and $\Lambda_1=\Lambda_2=\Lambda_3=(1-\kappa)/6$, along with an optimal unitary transform $V\in SU(6)$ such that
\begin{equation}
\begin{split}
\ket{0_V}_E&=\frac{1}{\sqrt{6}}\ket{0}_E-\frac{1}{\sqrt{6}}\ket{1}_E+\frac{1}{\sqrt{6}}\ket{2}_E+\frac{1}{\sqrt{6}}\ket{3}_E\\
  &+\frac{i}{\sqrt{6}}\ket{4}_E+\frac{1}{\sqrt{6}}\ket{5}_E,\\
  \ket{1_V}_E&=\frac{1}{\sqrt{2}}\ket{0}_E+\frac{1}{\sqrt{2}}\ket{1}_E,\\
  \ket{2_V}_E&=\frac{1}{\sqrt{2}}\ket{2}_E-\frac{1}{\sqrt{2}}\ket{3}_E,\\
  \ket{3_V}_E&=\frac{1}{\sqrt{2}}\ket{4}_E+\frac{i}{\sqrt{2}}\ket{5}_E,\\
  \ket{4_V}_E&=\frac{1}{2}\ket{0}_E-\frac{1}{2}\ket{1}_E-\frac{1}{2}\ket{2}_E-\frac{1}{2}\ket{3}_E,\\
  \ket{5_V}_E&=-\frac{1}{2\sqrt{3}}\ket{0}_E+\frac{1}{2\sqrt{3}}\ket{1}_E-\frac{1}{2\sqrt{3}}\ket{2}_E\\
  &-\frac{1}{2\sqrt{3}}\ket{3}_E+\frac{i}{\sqrt{3}}\ket{4}_E+\frac{1}{\sqrt{3}}\ket{5}_E,
\end{split}
\end{equation}
where $\ket{j}_E$ takes the computational basis as above. The last two vectors, $\ket{4_V}_E$ and $\ket{5_V}_E$, are irrelevant in computing $P^\star_E$, as the Schmidt rank of $\rho_{AB}$ is four, despite that all $\ket{k}_E$'s here must be six-dimensional.
The error rates equal to $\varepsilon_0=\varepsilon_1=\varepsilon_2=(1-\kappa)/3:=\varepsilon\in[0,~1/3]$ and the purification $\ket{\psi}_{ABE}=\sum_{k=0}^3 \sqrt{\Lambda_k}\ket{\phi_k}_{AB}\ket{k_V}_E$ are henceforth obtained. The maximum guessing probability equals
\begin{align}
  P^\star_E&=\sum_{j=0,1,2}\tr\biggr[\biggr(\Bigr|+n_j\Bigr\rangle\Bigr\langle +n_j\Bigr|_A\otimes\Bigr|2j\Bigr\rangle\Bigr\langle 2j\Bigr|_E
  +\nonumber\\
  &~~~~\Bigr|-n_j\Bigr\rangle\Bigr\langle -n_j\Bigr|_A\otimes\Bigr|2j+1\Bigr\rangle\Bigr\langle 2j+1\Bigr|_E \biggr)\Bigr|\psi\Bigr\rangle\Bigr\langle\psi\Bigr|_{ABE}\biggr]\nonumber  \\
  &=\frac{1}{2}+\sqrt{\frac{\Lambda_0}{3}}(\sqrt{\Lambda_1}+\sqrt{\Lambda_2}+\sqrt{\Lambda_3})\nonumber\\
  &=\frac{1}{2}+\sqrt{3\varepsilon(2-3\varepsilon)/4}.\label{pe-six}
\end{align}
It equals unity for $\varepsilon\in(1/3,~1/2]$.
We, again, plot the $P_B$-versus-$P_E$ figure (see Fig.~\ref{fig1}(b)) for arbitrary Bell-diagonal states, confirming the maximum of (\ref{pe-six}). It can be seen that (\ref{pe-six}) is lower than (\ref{pe-four}), since more observables impose more limits on Eve's guessing capability.

\begin{table}[tb]
\caption{\label{table}The critical values of error rates $\varepsilon_{\rm cr}$ and $\tilde\varepsilon_{\rm cr}$ around which $P_B\simeq P^\star_E$ and $\mathcal{R}\simeq0$, respectively, for the four-state protocol with measurements along the $\hat n_0=\hat z$ axis and an arbitrary direction $\hat n_1$ in the $\hat x \hat y$-plane. The parameters and variables are taken as $\theta_0=\varphi_0=0$, $\theta_1=\pi/2$, $\wp_0=\wp_1=1/2$, $\varepsilon_0=\varepsilon_1$, $\tilde\varepsilon_0=\tilde\varepsilon_1$, $\varepsilon=\sum_i \wp_i\varepsilon_i$, $\tilde\varepsilon=\sum_i \wp_i\tilde\varepsilon_i$, and $\delta\varepsilon:=\varepsilon_{\rm cr}-\tilde\varepsilon_{\rm cr}$. }
\begin{ruledtabular}
\begin{tabular}{cccccc}
 $\varphi_1$ & $0$  & $\pi/8$  & $\pi/4$ & $3\pi/8$ & $\pi/2$ \\
 \hline
 $\varepsilon_{\rm cr}~(\%)$    & $10.00$ & $11.06$ & $14.64$  & $11.06$ & $10.00$ \\
 $\tilde\varepsilon_{\rm cr}~(\%)$ & $11.00$  & $11.61$  & $12.62$ & $11.61$ & $11.00$ \\
 $\delta\varepsilon~(\%)$ & $-1.00$ & $-0.55$ & $+2.02$ & $-0.55$ & $-1.00$  \\
 $P^\star_E$ & $0.9000$ & $0.8894$ & $0.8536$ & $0.8894$ & $0.9000$
\end{tabular}
\end{ruledtabular}
\end{table}

Similar to the BB84 protocol, let $P_B>P^\star_E$ then we find $\varepsilon<(5-2\sqrt{3})/13\simeq11.8\%:=\varepsilon'_{\rm cr}$. The bound is still slightly smaller than that in the entropic criterion, since the same error rate yields $\mathcal{R}=1-h(3\varepsilon'_{\rm cr}/2)-(3\varepsilon'_{\rm cr}/2)\log_2 3\simeq0.045$. Note again that the information of the basis are $\log_2 3\simeq1.58$ bits, which have been deducted from $\mathcal{R}$. Also, let $\mathcal{R}=0$ then we find ${\tilde\varepsilon'}_{\rm cr}\simeq12.6\%$, yielding $P_B\simeq87.4\%$, which is slightly less than $P^\star_E\simeq89\%$.

\section{Summary and discussion}
We have presented a general analysis of quantifying the guessing probability in QKD protocols. In particular, we have proposed a simple guessing-probability-based relation, motivated by $\varepsilon_{\rm cor}$, and made comparison with the entropic security criteria by illustrating their varied determinations on the critical error rates in typical QKD protocols. Our results are in accord with the $\varepsilon$-security formalism according to which one must have $\varepsilon_{\rm cor}+\varepsilon_{\rm sec}\leq\varepsilon$. We have thus provided an alternative perspective on security issues. The results here may be useful to the QKD implementation.

The explicit relationship between the trace distance and the guessing probability, however, remains involved, because it depends on many aspects, such as eavesdropping strategies Eve takes to perform, specific QKD protocols Alice and Bob agree to use, various definitions of guessing probability, which phase of quantum cryptography the key considered is being processed in, etc. In many studies, guessing probability can be upper bounded by a function of the trace distance, as shown in~\cite{portmann14,portmann21} (where the definition is different from here). Debate was raised on whether the bound is a tight one~\cite{yuen16}. In \cite{li22}, researchers have shown that the bound can actually be reached in certain eavesdropping strategies, and that different eavesdropping strategies yield ``different numbers of guesses, sometimes even completely subversive differences, to get the final key.'' Thus the level in the $\varepsilon$-security should be carefully chosen, in order to guarantee the security of the keys. In our paper, what we have attained includes two main results: to upper bound the guessing probability via observables and to build an efficient guessing-probability-based relation applicable in security analysis. In the infinite-length limit, the bound is obtained in terms of error rates, instead of the trace distance, and the guessing-probability-based relation is rather efficient in determining the maximally tolerable error rates, as already shown in the previous sections. We shall consider finite-length cases in future studies. We expect that the finite-length guessing probability may also be upper bounded by some function of parameters in the $\varepsilon$-security.

The method we have used in this paper applies to any QKD protocols in which Alice's measuring settings are concurrent with Bob's specified in Sec.~\ref{symmetry}. It cannot directly apply to protocols using the Bell inequality~\cite{ekert91}, since Alice and Bob must perform Bell tests~\cite{bell64} and the Hilbert dimensions are usually unknown. To evaluate the guessing probability, a well received way is to redefine $P_B$ and $P_E$ as certain forms of Bell correlations. For instance, in~\cite{pawlowski10,hwang12,pawlowski12} where researchers consider security issues with a monogamous relation of nonlocal correlations, the guessing probability, or success probability, is a probabilistic form of the Bell-Clauser-Horne-Shimony-Holt inequality~\cite{chsh69}. The $P_B>P_E$ relation, however, cannot imply $\mathcal{R}>0$. In Ref.~\cite{masanes}, the device-independent QKD~\cite{ekert91,yao98,kent05,acin07} is investigated with various Bell inequalities. Their definition of the guessing probability is similar to ours (see Eq.~(18) in~\cite{masanes}) but bounded in terms of Bell violations. It is thus very worthwhile to study the guessing probability in device-independent protocols subsequently. Finally, due to noise and finite-length effects, the evaluation of probabilities may be non-ideal. One may resort to the law of large numbers to characterize the non-idealness. Suppose the ideal and actual distributions are denoted by $P^n$ and $Q^n$, respectively, where $n$ is the length of the generated key, and $P^n$ is an independent and identical distribution. Then, according to the law of large numbers, one can have ${\rm Pr}[~||P^n-Q^n||_1>\gamma~]\leq2^{-n\mathcal{D}}$~\cite{mcwilliams81}, where $||P^n-Q^n||_1:=\sum_x|P^n(x)-Q^n(x)|$ denotes the $\mathcal{L}_1$ distance, and $\mathcal{D}=\frac{\gamma^2}{2\ln2}-d\times\frac{\log_2(n+1)}{n}$. Equivalently, letting $2^{-n\mathcal{D}}:=\epsilon$, one gets $\gamma=\left(\frac{2\ln2}{n}\left( \log_2\frac{1}{\epsilon}-d\times\log_2\frac{1}{n+1} \right)\right)^{1/2}$. That is, the probability that the actual distribution deviates from the ideal one with a degree larger than $\gamma$ is no larger than $\epsilon$.

\acknowledgments
The study is supported by the National Natural Science Foundation of China and the NSAF.

\end{document}